\documentclass[conference,10pt]{IEEEtran}
\IEEEoverridecommandlockouts
\usepackage{cite}
\usepackage{amsmath,amssymb,amsfonts}
\usepackage{algorithmic}
\usepackage{graphicx}
\usepackage{textcomp}
\usepackage{xcolor}
\usepackage{booktabs}
\usepackage{tabularx}
\usepackage{multirow}
\usepackage{hyperref}
\usepackage{listings}
\usepackage{cancel}
\hypersetup{
    colorlinks=true ,
    linkcolor=black,
    citecolor=black }
\def\BibTeX{{\rm B\kern-.05em{\sc i\kern-.025em b}\kern-.08em
    T\kern-.1667em\lower.7ex\hbox{E}\kern-.125emX}}
\begin{document}

\title{From Ranking to Reasoning: Explainable Web API Recommendation via Semantic Reasoning  
}

\author{\IEEEauthorblockN{Zishuo Xu\IEEEauthorrefmark{1}, Dezhong Yao\IEEEauthorrefmark{2}, Yao Wan\IEEEauthorrefmark{2}}
\IEEEauthorblockA{
\IEEEauthorrefmark{1}\textit{School of Software Engineering} 
\IEEEauthorrefmark{2}\textit{School of Computer Science and Technology} \\
\textit{Huazhong University of Science and Technology}\\
Wuhan, Hubei, China \\
Email: ZishuoXuHUST@163.com, \{dyao, wanyao\}@hust.edu.cn}
}

\maketitle
\setcounter{page}{1}
\begin{abstract}
The rapid growth of Web APIs has made automated Web API recommendation essential for efficient mashup development.
However, existing approaches suffer from two major limitations: 1) they rely on fixed top-$N$ recommendation strategies that cannot adapt to mashup complexity, and 2) they provide little or no explanation for recommended APIs, limiting transparency and user trust.
To address these challenges, we propose WAR-R1, an explainable 
Web API recommendation framework that integrates semantic reasoning with adaptive, variable-cardinality recommendation. Built on a lightweight large language model (LLM), WAR-R1 generates both a set of relevant APIs and a natural-language justification for each recommendation.
To support adaptive recommendation size, we introduce special start and stop tokens that allow the model to learn when to begin and terminate API generation. WAR-R1 is trained in two stages: supervised fine-tuning on an annotated mashup–API corpus, followed by reinforcement learning using Group Relative Policy Optimization (GRPO) with low-rank adaptation to jointly optimize recommendation accuracy and reasoning quality. 
Experiments on the ProgrammableWeb dataset show that WAR-R1 outperforms state-of-the-art baselines by up to 10.89\% in recommendation accuracy while consistently producing high-quality, semantically grounded explanations.  Extensive ablation studies validate the effectiveness of reinforcement learning, special token design, and integrated reasoning.
\end{abstract}


\section{Introduction}\label{sec:introduction}

With the rapid advancement of cloud computing and service-oriented architectures, the number of publicly available Web APIs has grown dramatically~\cite{10045831, 10912769}. 
For example, ProgrammableWeb reported more than 24,000 registered APIs in 2022, representing over a tenfold increase in the past decade.
Web APIs enable developers to reuse remote functionality and data through lightweight interfaces, forming the foundation of modern Web and mobile applications. This growth has also driven the widespread adoption of mashups, which integrate Web APIs without building services from scratch~\cite{10542468,gong2022dawar}.

Despite their advantages, mashup development faces a fundamental challenge: identifying an appropriate set of APIs that collectively satisfy a mashup’s functional requirements. As the API ecosystem continues to expand, developers—especially those without extensive domain expertise—must evaluate a large and heterogeneous set of candidate APIs, leading to high development overhead and suboptimal service selection. Thus, automated Web API recommendation has become a critical research problem, aiming to guide developers toward relevant, compatible APIs and reduce integration effort~\cite{wang2023functional,WangXLPWDY24}.
 
\begin{figure}[t]
\centering
\includegraphics[width=\linewidth]{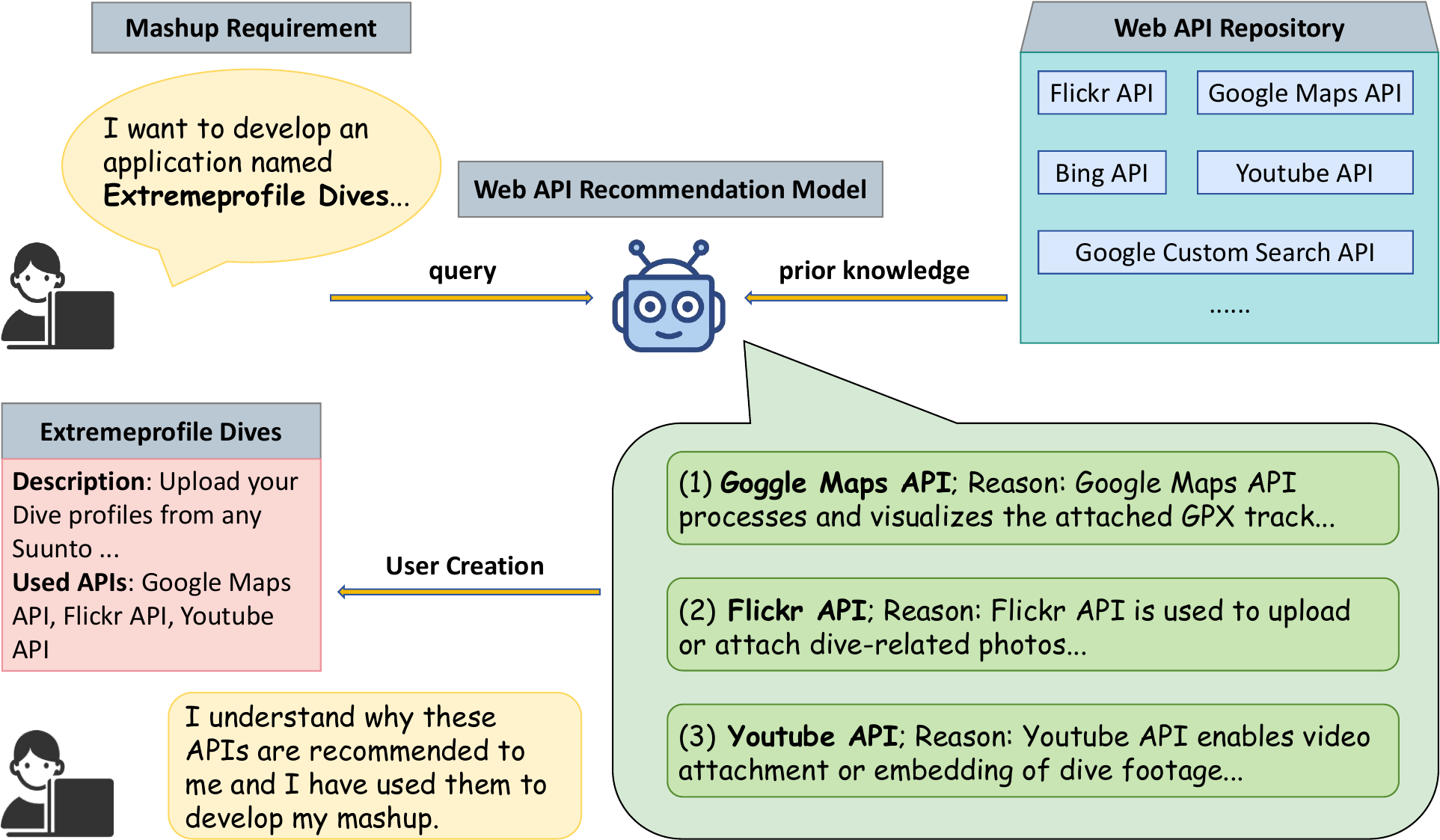}
\vspace{-1.8em}
\caption{Example of Web API recommendation via semantic reasoning.}
 \label{fig:ar}
\vspace{-1.5em}
\end{figure}

Over the past decade, numerous Web API recommendation approaches have been proposed, including content-based methods \cite{10045831,icsocWangZWLLW21}, collaborative filtering (CF) \cite{Chen2014QoSAwareWS,YaoWSBH21}, and hybrid strategies \cite{wang2023functional,WangXLPWDY24}. While these methods have demonstrated encouraging results, two fundamental limitations remain unresolved. 
1) Most existing approaches rely on fixed top-$N$ recommendation strategies, returning a predetermined number of APIs for every mashup. This uniform strategy ignores the fact that mashups vary significantly in complexity: simple mashups may require only one or two APIs, whereas complex applications may depend on many more. As a result, fixed-cardinality recommendations often introduce irrelevant APIs for simple mashups or omit essential services for complex ones, reducing recommendation effectiveness and practical usability.
2) Lack of explanation for the user. Most methods output a ranked list of APIs without providing reasons for their inclusion. This opacity prevents developers from understanding how recommendations relate to their requirements, making it difficult to assess correctness, or debug errors. In real-world mashup development, developers often need to understand why an API is recommended in order to decide whether it fits functional, architectural, or integration constraints \cite{10.1016/j.ins.2025.122049,wang2023deep}.

Recent advances in large language models (LLMs) have shown strong capability in understanding natural language requirements \cite{ouyang2022training,10660826} and generating coherent explanations \cite{wei2022chain}, opening new opportunities for semantic-aware recommendation. Several recent works have leveraged LLMs to enhance API representation and recommendation accuracy \cite{ZOU2025104219,10948476}. However, these approaches still primarily focus on ranking accuracy, typically adopt fixed top-$N$ outputs, and do not provide explicit, per-API explanations grounded in mashup requirements and API functionality.

To tackle these challenges, we propose WAR-R1\footnote{The source code is available at \url{https://github.com/ZsXu-enrico/WAR-R1}}, an Explainable  \underline{W}eb \underline{A}PI  \underline{R}ecommendation model via Semantic \underline{R}easoning.
WAR-R1 is an LLM-based model that jointly performs Web API recommendation and explanation generation.
Unlike prior approaches, WAR-R1 supports variable-cardinality recommendation, allowing the model to adaptively determine how many APIs to recommend based on mashup complexity. This is achieved by introducing task-specific start and stop tokens that explicitly control the generation boundaries of recommended APIs.
Moreover, WAR-R1 provides natural-language justifications for each recommended API, enabling developers to understand how individual APIs satisfy specific aspects of mashup requirements.
Figure \ref{fig:ar} illustrates this framework using the “Extremeprofile Dives” mashup as an example. Specifically, we begin by enriching our training corpus using DeepSeek‑R1 \cite{deepseekai2025}, which annotates each example with a natural language rationale linking the mashup description to the target API’s functionality. WAR‑R1 is then fine‑tuned in two phases: 1) supervised fine-tuning on the annotated mashup–API dataset enriched with semantic rationales. 
2) reinforcement learning via the Group Relative Policy Optimization (GRPO)  \cite{shao2024deepseek} with Low-Rank Adaptation (LoRA) \cite{hu2021loralowrankadaptationlarge} to jointly optimize recommendation accuracy and reasoning quality.
This reinforcement learning stage enables WAR-R1 to explore diverse API combinations and align its recommendations with both quantitative performance metrics and explanation consistency.
We evaluate WAR-R1 on the ProgrammableWeb dataset and compare it against SOTA baselines. WAR-R1 outperforms existing methods in both accuracy and semantic explanation quality.
The main contributions of this work are summarized as follows:
\begin{itemize}
    \item 
We propose WAR-R1, a novel LLM-based Web API recommendation framework that integrates semantic reasoning and supports adaptive, variable-cardinality recommendation.
    \item 
We introduce a two-stage training paradigm leveraging GRPO-based 
reinforcement learning with multi-objective reward signals to jointly 
optimize recommendation accuracy and explanation quality.
    \item We conduct comprehensive experiments and ablation studies demonstrating that WAR-R1 outperforms SOTA baselines and produces interpretable, developer-oriented API recommendations.

\end{itemize} 
\section{Related Work}
\label{sec:relwork}
\subsection{Web API Recommendation}
Existing Web API recommendation approaches can be categorized into: content-based methods, collaborative filtering (CF) methods, hybrid models, and LLM-based approaches.
\textbf{Content-based methods} recommend APIs by matching mashup requirements with API descriptions \cite{CaoPZQTKL23}. Shi et al. \cite{shi2019functional} apply attention-enhanced LSTM models to identify salient segments in mashup descriptions. ServiceBERT \cite{icsocWangZWLLW21} leverages BERT to encode service semantics. Bao et al. \cite{8029765} formulate API link prediction to enhance mashup recommendations. Sang et al. \cite{10045831} employ full-text semantic mining of developer requirements. 
\textbf{Collaborative filtering methods} leverage mashup–API invocation data, assuming that similar mashups tend to use similar APIs.
Chen et al. \cite{Chen2014QoSAwareWS} propose a hybrid CF approach that combines user-based and item-based strategies, incorporating QoS metrics to assess API suitability. Yao et al. \cite{YaoWSBH21} develop a probabilistic matrix factorization model that accounts for explicit similarity measures and latent API correlations to optimize mashup co-invocation predictions. Wang et al. \cite{wang2021novel} integrate knowledge graph embeddings with CF, and model semantic relationships between APIs and mashups. 
To overcome the limitations of individual paradigms, \textbf{hybrid methods} combine multiple information sources.
Xia et al. \cite{XiaFTHZW15} propose a category-aware clustering of services combined with CF to predict category-specific rankings. Qi et al. \cite{9763334} construct a correlation graph to generate tailored API suggestions that meet both functional requirements and compatibility constraints. Wang et al. \cite{wang2023functional} develop FSFM, fusing requirement semantic vectors with representations of structural features. SEHGN \cite{WangXLPWDY24} uses heterogeneous graph neural networks and semantic embeddings to model both the links between mashups and APIs and their text descriptions. Recent \textbf{LLM-based approaches} enable more expressive semantic understanding and enhanced API representation.
LLMSRec~\cite{PENG2025113710} augments LLMs with service networks to bridge semantic gaps, while SRCA~\cite{ZOU2025104219} improves service representations using category co-occurrence graphs.
Qin et al. \cite{10948476} present a framework for precise mashup API suggestions via multitask fine-tuning. 

However, all the above methods still primarily focus on ranking performance, typically output fixed-size recommendation lists, and do not generate explicit, per-API explanations grounded in mashup requirements and API functionality.
\subsection{Reinforcement Learning for Large Language Models}
Reinforcement learning (RL) has recently played an important role in improving LLM reasoning and alignment~\cite{ouyang2022training}.
Methods such as Direct Preference Optimization (DPO)~\cite{rafailov2023direct} and Group Relative Policy Optimization (GRPO)~\cite{shao2024deepseek} have been applied to enhance reasoning quality without requiring explicit critic models. State-of-the-art reasoning-focused LLMs, including DeepSeek-R1, Qwen2.5~\cite{qwen2025qwen25technicalreport}, and Kimi k1.5~\cite{kimiteam2025}, demonstrate that RL can significantly improve structured generation and decision-making capabilities. Seed1.5‑Thinking \cite{seed2025seed15} augments its RL toolkit with Value-model-based Augmented Proximal Policy Optimization (VAPO) and Dynamic sAmpling Policy Optimization (DAPO), leveraging these frameworks to stabilize training and enhance reasoning capabilities across complex tasks.

Inspired by these advances, we adopt GRPO to fine-tune WAR-R1 after supervised training. Unlike prior Web API recommendation studies, which rely solely on supervised learning, our approach leverages online reinforcement learning to jointly optimize recommendation accuracy and explanation consistency. To the best of our knowledge, WAR-R1 is the first Web API recommendation framework to integrate GRPO-based reinforcement learning with explicit generation control for adaptive recommendation and semantic reasoning.

\section{Problem Definition and Task Formulation}
\label{sec:prodef}

We formalize Web API recommendation with semantic reasoning as a joint prediction task. The goal is to recommend a set of Web APIs for a given mashup and provide an explicit justification for each recommendation.
Let \(m_i\) denote a mashup with requirement description \(T^{m_i}\), as 
specified by the user. Let
$A = \{a_1, \ldots, a_j, \ldots, a_S\}$
represent a repository comprising \(S\) Web APIs. Each API $a_j = \langle D^{a_j}, C^{a_j}, X^{a_j}  \rangle \in A$ is defined by its functional description $D^{a_j}$, a category $C^{a_j}$, and associated meta‐elements $X^{a_j}$. Given \(T^{m_i}\) and \(A\), the task is twofold: 
\begin{enumerate}
  \item \textbf{Adaptive API Recommendation:} Predict a subset $S^{m_i} = \{a_1, \ldots, a_l, \ldots, a_s\} \subseteq A$ of APIs that best satisfy the mashup requirements. Unlike traditional approaches that return a fixed top-$N$ list, the size of $S^{m_i}$ is a dynamic $s$ and not predefined.
  \item \textbf{Semantic Reasoning:} For each recommended API \(a_l\)$\in S^{m_i}$, generate a natural-language explanation \(r_{a_l}^{m_i}\) that justifies why the API is suitable for the mashup based on the mashup description and API functionality.
\end{enumerate}

\noindent The output for mashup  $m_i$ is therefore a set of API-reason pairs:
$\{(a_l, r^{m_i}_{a_l}) \mid a_l \in {S}^{m_i}\}.$ 
In the ideal scenario, the recommended API set \(S^{m_i}\) matches the APIs actually used by developers in the mashup: $A^{m_i}=S^{m_i}$. This formulation generalizes traditional Web API recommendation by jointly modeling what APIs to recommend, how many APIs to recommend, and why each API is recommended.

\section{Methodology}
\label{sec:method}
\subsection{Overview of WAR-R1}
\begin{figure*}[t]
\centering
\includegraphics[width=0.9\linewidth]{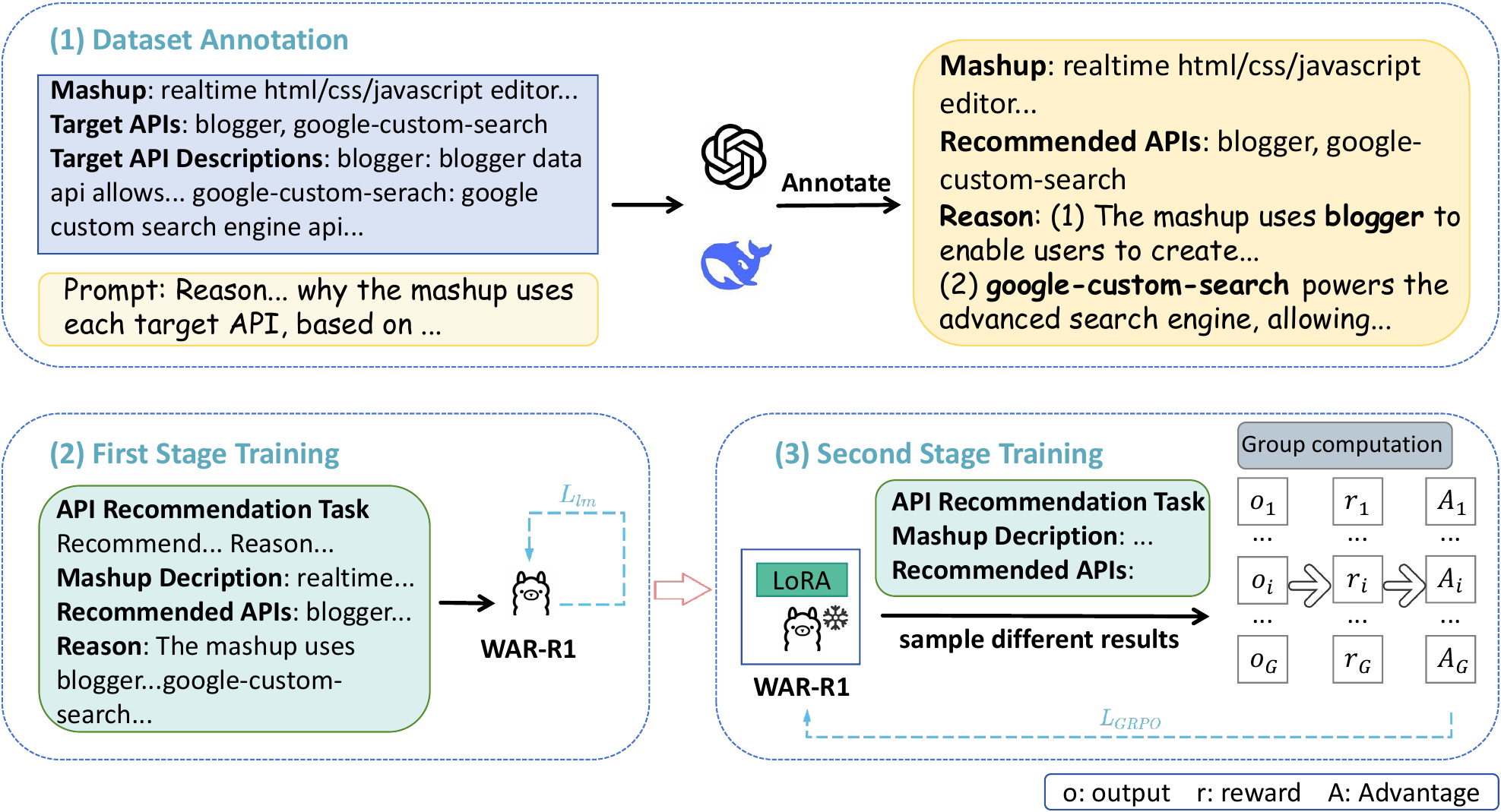}
\vspace{-1.0em}
\caption{The architecture of WAR-R1.}
\label{fig:arc}
\vspace{-1.0em}
\end{figure*}

Figure~\ref{fig:arc} illustrates the overall architecture of WAR-R1, an LLM-based framework designed to jointly perform adaptive Web API recommendation and semantic reasoning. WAR-R1 is built on the TinyLlama \cite{zhang2024tinyllama} backbone and is trained in two stages: supervised fine-tuning and reinforcement learning. The framework consists of three main components: 1) \textbf{Dataset annotation} with natural-language rationales, 2) \textbf{Structured generation} using task-specific control tokens, and 3) \textbf{Reinforcement learning optimization} to align recommendations and explanations.

For the Supervised fine-tuning, we perform standard language modeling with a cross-entropy loss to train the model to predict the next token in a sequence of mashup descriptions, recommended APIs and rationale texts. After supervised fine-tuning, WAR-R1 acquires foundational capabilities for API recommendation and semantic reasoning. For the GRPO Reinforcement Learning with Low-Rank Adaptation (LoRA), We then fine-tune the model via GRPO, incorporating LoRA to restrict trainable parameters to a lightweight subset, thereby reducing computational cost. After reinforcement learning, WAR-R1 further enhances the performance, especially in alignment between recommendation and reasoning by ensuring more recommended APIs are accompanied by justificatory reasons.

\subsection{Dataset Annotation with Semantic Rationales}
To train WAR-R1, we need a dataset that pairs historical mashup-API invocations with high-quality explanatory text. 
However, historical mashup–API datasets typically include only mashup descriptions and the APIs used, without explicit explanations.
We address this by annotating the dataset with LLM-generated explanations. For each mashup, the annotation prompt includes: the mashup requirement description, the target APIs, and the official functional descriptions of those APIs.
Our structured prompt is given in Table~\ref{tab:prompt-reason}. This design grounds the generated rationales in concrete API functionality, reducing hallucinations and encouraging explanations that explicitly connect mashup requirements with API capabilities.
\setlength{\tabcolsep}{1mm}

\begin{table}[!t]
  \centering

    \caption{The prompt and example answer for dataset annotation.}
  \begin{tabular}{@{}p{0.18\columnwidth}|p{0.77\columnwidth}|@{}}
    \toprule
    \textbf{Prompt} 
      & \textbf{API Recommendation Reasoning Task}  
        
        Reason in English, concisely and accurately explain why the mashup uses each target API, based on the descriptions of the mashup and the target APIs.  
        When you mention an API name, wrap it with three asterisks exactly as given in Target APIs (e.g., ***sendgrid***).  
        
        \textbf{Mashup}: realtime html/css/javascript editor and advanced search engine dedicated to web development and design.  
        
        \textbf{Target APIs}: blogger, google-custom-search  
        
        \textbf{Target API Descriptions}: blogger: blogger data api allows client application to view update blogger content...; google-custom-search: google custom search engine api restful api allows developer get web image search result...
    \\\midrule
    \textbf{Answer} 
      & The mashup uses ***blogger*** to enable users to create, manage, and publish web development tutorials, code snippets, or design blogs directly within the editor, leveraging the API's ability to programmatically handle blog content. ***google-custom-search*** powers the advanced search engine, allowing users to query web development resources, documentation, or community solutions 
      in real-time, enhancing the editor's utility. 
    \\\bottomrule
  \end{tabular}

  \label{tab:prompt-reason}
\end{table}

\subsection{Stage 1: Supervised Fine-Tuning}
Building on the annotated dataset, we conduct supervised fine-tuning to equip WAR-R1 with foundational capabilities for API recommendation and semantic reasoning.
Since most API names are initially out of the tokenizer's vocabulary, we explicitly add each API as a single indivisible token (e.g., \texttt{<API\_sendgrid>}) to prevent the tokenizer from decomposing an API name into multiple tokens and to enable the model to treat each API as a distinct, atomic concept.
We introduce task-specific delimiters to structure the model’s output: \texttt{<API\_start>} and \texttt{<API\_stop>} frame the API recommendation sequence, while \texttt{<REASON\_start>} and \texttt{<REASON\_stop>} enclose the explanatory reasoning. This design enables the model to clearly distinguish between the two subtasks and learn precise generation boundaries. During training, the input sequence is constructed according to the template in Table~\ref{tab:prompt}. The template includes three key parts: the mashup's requirement description; a recommended API list enclosed by \texttt{<API\_start>} and \texttt{<API\_stop>}; explanatory reasons for each API, wrapped by \texttt{<REASON\_start>} and \texttt{<REASON\_stop>}. The mashup descriptions and recommended APIs are derived from historical records, and the explanatory reasons are annotated by an LLM. The model learns patterns in mashup descriptions to improve recommendations for similar mashups.

\setlength{\tabcolsep}{1mm}
\begin{table}[!t]
  \centering
  \caption{The prompt for the two-stage training of WAR-R1.}
  \begin{tabular}{@{}p{0.18\columnwidth}|p{0.77\columnwidth}|@{}}
    \toprule
    \textbf{First Stage Prompt}
      & \textbf{API Recommendation Task}  
        
        Recommend APIs for the mashup according to its description and give the reason for each recommendation.  
        
        \textbf{Mashup Description}:  
        realtime html/css/javascript editor and advanced search engine dedicated to web development and design.  
        
             \textbf{Recommended APIs}: \texttt{<API\_start>}
        
         \texttt{<API\_blogger> <API\_google-custom-search> <API\_stop>}  
        
        \textbf{Reason}:  
        \texttt{<REASON\_start>}  
        The mashup utilizes \texttt{<API\_blogger>} to enable users to create, manage, and publish web ... 
        
        \texttt{<API\_google-custom-search>} powers the advanced search engine, allowing users to query web development ...\texttt{<REASON\_stop>}
    \\\midrule
          \textbf{Second Stage Prompt}
        & \textbf{API Recommendation Task}  
          
          Recommend APIs for the mashup...  
          
          \textbf{Mashup Description}:  
          realtime html...  
          
          \textbf{Recommended APIs}:  
      \\\bottomrule

  \end{tabular}

  \label{tab:prompt}
\end{table}

Let $H$ denote the total length of the target token sequence and $w_{h}$ the $h$-th token. The model parameterized by $\Phi$ produces a predictive distribution $P_{\Phi}(w_{h}\mid w_{1},\dots, w_{h-1})$ at each step. We adopt the standard cross-entropy loss over the entire sequence:

\begin{align}\mathcal{L}_\Phi = -\frac{1}{H} \sum\limits_{h=1}^H \log P_\Phi(w_h \mid w_1, \ldots, w_{h-1}).\end{align}
Here, the term $\log P_\Phi(w_h \mid w_1, \ldots, w_{h-1})$ measures the log-likelihood of the true token \(w_h\) under the model’s predictive distribution. Negating and averaging these log-likelihoods across all positions $h$ in the sequence encourages the model to prioritize correct tokens at each step.
This formulation ensures the model learns foundational capabilities for generating syntactically valid sequences that respect the structure imposed by special tokens (e.g., transitioning from \(\texttt{<API\_start>}\) to API tokens to \(\texttt{<API\_stop>}\)), selecting APIs relevant to the mashup description, and producing coherent reasons that align with the recommended APIs.

\subsection{Stage 2: Reinforcement Learning with GRPO}
\subsubsection{Overall Introduction}
The first stage training equips WAR-R1 with foundational capabilities for 
generating well-structured outputs. However, supervised fine-tuning is inherently limited by its reliance on static training data. To enable exploration beyond the historical patterns, we adopt GRPO for the second stage. Unlike traditional policy optimization methods that require a separate critic network, GRPO samples multiple outputs for each input and computes advantages through group-relative comparisons, allowing the model to discover API combinations that better satisfy evaluation metrics. For each input question \(q\) omitting ground-truth APIs and reasons as given in Table~\ref{tab:prompt}, GRPO samples a group of outputs \(\{o_1,o_2,\dots,o_G\}\) from the old policy \(\pi_{\theta_{\mathrm{old}}}\) and then optimizes the policy model by maximizing the following objective:

\begin{multline}
\mathcal{J}_{\mathrm{GRPO}}(\theta)
=
\mathbb{E}_{\substack{q \sim P(Q),\\ \{o_i\}\sim\pi_{\theta_{\mathrm{old}}}}}
\Bigl[
\frac{1}{G}\sum_{i=1}^G\frac{1}{|o_i|}
\sum_{t=1}^{|o_i|} \\
\min\Bigl\{
\frac{\pi_\theta(o_{i,t}\mid q,o_{i,<t})}
     {\pi_{\theta_{\mathrm{old}}}(o_{i,t}\mid q,o_{i,<t})}\,\hat A_{i,t},\,
\mathrm{clip}\bigl(\frac{\pi_\theta}{\pi_{\theta_{\mathrm{old}}}},1-\varepsilon,1+\varepsilon\bigr)\,\hat A_{i,t}
\Bigr\} \\
-\,\beta\,D_{\mathrm{KL}}\bigl[\pi_\theta\|\pi_{\mathrm{ref}}\bigr]
\Bigr].
\end{multline}

Here \(\varepsilon\) and \(\beta\) are hyper‐parameters, and \(\hat A_{i,t}\) is the advantage computed using a group of rewards \( \{r_1, r_2, \dots, r_G\} \) corresponding to the outputs within each group:
\begin{align}
\hat A_{i,t} = \frac{r_i - \text{mean}\bigl(\{r_1, r_2, \cdots, r_G\}\bigr)}{\text{std}\bigl(\{r_1, r_2, \cdots, r_G\}\bigr)}.
\end{align}
GRPO incorporates regularization by adding the KL divergence between the learned policy and the reference policy directly into the loss, thereby sidestepping any added complexity in computing \(\hat A_{i,t}\). In our implementation, the reference policy \(\pi_{\mathrm{ref}}\) is set to the old policy \(\pi_{\theta_{\mathrm{old}}}\), ensuring alignment with the policy update framework.

To reduce computational cost, we adopt LoRA in our GRPO training. For a given weight matrix \(\mathbf{W}_0 \in \mathbb{R}^{d \times k}\), LoRA employs a low-rank decomposition to modify the weights, expressed as \(\mathbf{W}_0 + \Delta \mathbf{W} = \mathbf{W}_0 + \mathbf{B}\mathbf{A}\), where \(\mathbf{B} \in \mathbb{R}^{d \times r}\) and \(\mathbf{A} \in \mathbb{R}^{r \times k}\), with the rank \(r \ll \min(d, k)\). During the training process, \(\mathbf{W}_0\) remains fixed and does not receive gradient updates, while the matrices \(\mathbf{A}\) and \(\mathbf{B}\) are trainable. Thus, in our GRPO stage we only optimize the LoRA parameters $\{\mathbf{A},\mathbf{B}\}$.
\subsubsection{Design of the Reward Function}
Our overall reward consists of two components: (i) Recommendation quality and (ii) Reasoning quality.

For the recommendation reward, we employ standard information retrieval metrics to assess API recommendation performance. As $s$ denotes the number of recommended APIs output by the model, we use “@$s$” to indicate that these metrics are computed over all APIs recommended by the model.
Precision@$s$ refers to the ratio of the number of real hits in the recommended APIs to the number of recommended APIs. 
\begin{align}
    \text{Precision}@s = \frac{|\text{{APIs}}_\text{{real}} \cap \text{APIs}_\text{recommend }|}{| \text{APIs}_\text{recommend}|}.
\end{align}
Recall@$s$ is defined as the ratio of the number of real hits in the recommended APIs to the number of all real APIs. 
\begin{align}
    \text{Recall}@s = \frac{|\text{APIs}_\text{real} \cap \text{APIs}_\text{recommend}|}{|\text{APIs}_\text{real}|}.
\end{align}
Furthermore, the F1 score, defined as the harmonic mean of Precision and Recall, is given by:
\begin{align}
\text{F}1 = \frac{2 * \text{Precision} * \text{Recall}}{\text{Precision} + \text{Recall}}\,.
\end{align}

Since Precision@$s$ and Recall@$s$ do not take into account the order in the results, we use additional evaluation metrics: Mean Average Precision (MAP) and Normalized Discounted Cumulative Gain (NDCG). NDCG is calculated from DCG by normalizing IDCG which is the ideal DCG. 
NDCG@$s$ is given by:
\begin{gather}
    \text{DCG}@s = \sum_{i=1}^s\frac{rel_i}{\log_2(i + 1)}, \\
    \text{IDCG}@s = \sum_{i=1}^{|\text{real APIs}|}\frac{1}{\log_2(i + 1)}, \\
    \text{NDCG}@s = \frac{\text{DCG}@s}{\text{IDCG}@s}.
\end{gather}
where $rel_i \in \{0,1\}$ represents whether the $i$-th API is truly relevant to the current mashup or not.

Average Precision (AP) calculates Precision at each position, and averages these values over all matching positions.
MAP is obtained by averaging AP across all mashups.
In practice, the batch size is set to 1 during the second‐stage training. As a result, the reward function at this stage directly uses AP@$s$ computed for each individual mashup.
AP@$s$ is given by:
\begin{align}
    \text{AP}@s = \frac{\sum_{i=1}^s rel_i * \text{Precision}@i}{\sum_{i=1}^s rel_i}.
\end{align}
MAP@$s$ is given by:
\begin{align}
    \text{MAP}@s = \frac{1}{|M|} \sum_{m \in M} \text{AP}_m@s.
\end{align}
where $M$ is the set of all mashups.
The recommendation reward is given as:
\begin{equation}
\begin{split}
\text{recommendation reward} = 0.4*\text{F1} + 0.1*\text{Precision}\\+ 0.1*\text{Recall}+ 0.2*\text{NDCG}+ 0.2*\text{AP}.
\end{split}
\end{equation}
Since F1 already balances Precision and Recall, we assign it the highest weight. However, we also retain small individual weights for Precision and Recall to additionally reward outputs that excel in either metric.

For the reasoning reward, we introduce Reasoning Precision (RP) and Reasoning Recall (RR) to evaluate the quality of reasoning. RP quantifies the proportion of all provided reasons that correspond to APIs actually recommended by the model.
\begin{align}
    \text{RP} = \frac{|\text{{APIs}}_\text{{recommend}} \cap \text{APIs}_\text{reason }|}{| \text{APIs}_\text{reason}|}.
\end{align}
RR quantifies the proportion of all APIs recommended by the model that are accompanied by a valid reason.
\begin{align}
    \text{RR} = \frac{|\text{{APIs}}_\text{{recommend}} \cap \text{APIs}_\text{reason}|}{| \text{APIs}_\text{recommend}|}.
\end{align}
RP and RR are equally weighted, so the reasoning reward is formulated as:
\begin{equation}
\begin{split}
\text{reasoning reward} = 0.5*\text{RP} + 0.5*\text{RR}.
\end{split}
\end{equation}

The overall reward is the sum of the recommendation reward and reasoning reward, as both components carry equal significance in our model design.

\section{Experiments And Analysis}
\label{sec:expanalysis}
To evaluate the effectiveness of WAR-R1, we need to answer the following questions:

\noindent RQ1: How does WAR‑R1 perform compared to state-of-the-art Web API recommendation methods?

\noindent RQ2: How effective is WAR-R1 in semantic reasoning?

\noindent RQ3: What is the impact of the reinforcement learning stage on WAR‑R1’s overall performance?

\noindent RQ4: How do the special control tokens affect performance?

\noindent RQ5: Does integrating reasoning improve recommendation quality?

\noindent RQ6: What are the computational costs of WAR-R1?

\subsection{Dataset and Experimental Settings}
 We conduct experiments on the widely used dataset ProgrammableWeb~\cite{wu2021mashup}, which contains 8,217 mashups and 1,647 Web APIs. Detailed statistics of the dataset are presented in Table~\ref{tab:dataset}. 
 Each mashup is associated with a textual description and a set of invoked APIs. Following common practice, we randomly split the mashups into training, validation, and test sets with a ratio of 3:1:1, while keeping the full API repository available during testing. WAR-R1 is built on the TinyLlama backbone. The supervised fine-tuning stage is trained for 20 epochs with a learning rate of 1e-5, and the reinforcement learning stage is trained for 10 epochs with a learning rate of 5e-6. All experiments are conducted on four NVIDIA V100 GPUs (32GB).

\setlength{\tabcolsep}{1mm}
\begin{table}[t]
\centering
\caption{Statistics of the dataset.}
\begin{tabular}{l|c}
\toprule
\textbf{Statistics}& \textbf{Value}\\
\midrule
Number of APIs& 1647 \\
Number of Mashups& 8217\\
Number of Categories& 499\\
APIs per Mashup& 2.091\\
Words in description per Mashup& 17.019\\
Words in description per API& 44.296\\
\bottomrule
\end{tabular}
\label{tab:dataset}
\vspace{-1.0em}
\end{table}

\subsection{Baseline Models}
We compare WAR‑R1 against eight representative baselines including 1) zero-shot LLMs: Llama-3.3-70B, Qwen3-235B-A22B, GPT-4.1; 2) deep learning methods: MTFM, CLJT, BERT-CM, SEHGN; 3) LLM-based approach: LLMAR.

\begin{itemize}

    \item \textbf{Llama-3.3-70B} \cite{grattafiori2024llama3herdmodels}: Llama-3.3-70B is an instruction-tuned multilingual large language model with 70 billion parameters.
    \item \textbf{Qwen3-235B-A22B}~\cite{yang2025qwen3technicalreport}: Qwen3-235B-A22B is a mixture-of-experts large language model with 235 billion total parameters.
    \item \textbf{GPT-4.1}: GPT-4.1 is an advanced large language model from OpenAI, known for its strong instruction-following and reasoning capabilities.
    \item \textbf{MTFM} \cite{wu2021mashup}: MTFM uses a multi-layer CNN to embed mashups and APIs, concatenates their embeddings for collaborative filtering, and employs multi-task learning with mashup category prediction.
    \item \textbf{CLJT}~\cite{10793452}: CLJT explores textual and structural semantics using contrastive learning to derive discriminative feature representations and employs joint training of representation and recommendation tasks.
       \item \textbf{BERT-CM} \cite{10489941}: BERT-CM uses BERT to encode mashup descriptions and is trained through a linear classification layer.
        \item \textbf{SEHGN} \cite{WangXLPWDY24}: SEHGN combines heterogeneous graph neural networks with semantic embedding techniques to model complex structural dependencies in mashup-API interaction networks and  textual descriptions.
    \item \textbf{LLMAR} \cite{10948476}: LLMAR fine-tunes Llama-3.2-3B on mashup and API categorization, API description, and API recommendation tasks.
\end{itemize}

\subsection{Evaluation Metrics}
\noindent\textbf{Recommendation Performance:} We evaluate recommendation accuracy using standard information retrieval metrics:
Precision@\(s\) (P@$s$), Recall@\(s\) (R@$s$), NDCG@\(s\)  (N@$s$), and MAP@\(s\) (M@$s$). Unlike many prior works that fix \(N\) to evaluate Top-\(N\) recommendations, our model outputs an average between 1.7 and 2.6 APIs per query depending on the temperature. Therefore, for baseline methods we report metrics at \(N = 1\) and \(N = 3\), while for our approach we evaluate at \(s = 1\) and at a dynamically determined cutoff \(s\). Higher values of these metrics correspond to better recommendation accuracy and ranking quality.

\noindent\textbf{Reasoning Quality:} To assess explanation quality, we employ
Reasoning Precision (RP) and Reasoning Recall (RR), which measure the alignment between recommended APIs and generated reasons. In addition, we introduce Reasoning Score (RS) evaluated by the LLM.  We employ GPT-4.1 as the judge due to its widespread adoption in LLM-as-a-judge paradigms and its demonstrated alignment with human preferences \cite{zheng2023judging,liu-etal-2023-g}. This choice also mitigates potential evaluation bias, as using DeepSeek-R1 for evaluation might favor its own expression style, given that our model is fine-tuned on DeepSeek-R1's annotations. Specifically, we provide GPT-4.1 with the mashup description, the set of recommended APIs and their function descriptions, and the generated reasons, then ask it to assign a score in \([0,1]\) based on three criteria independently:  
  (1) Mashup Alignment: The degree to which the reason addresses the mashup requirements;
  (2) API Alignment: The consistency of the reason with the official API descriptions;
  (3) Readability: The overall clarity and readability of the explanation.
The exact prompt template and an example response are shown in Table~\ref{tab:prompt-score}.

\setlength{\tabcolsep}{1mm}
\begin{table}[!t]
\centering
\caption{The prompt and example answer for Reasoning Score.}
\label{tab:prompt-score}
  \begin{tabular}{@{}p{0.18\columnwidth}|p{0.78\columnwidth}|@{}}
 
      \toprule
      \textbf{Prompt} 
        & \textbf{API Reason Evaluation Task}

You are given a mashup description, recommended APIs, their corresponding descriptions, and the reason for the recommendation.

\textbf{\#Instruction}

Please evaluate the recommendation by:

\textbf{1.} List all the APIs both in the “Recommended APIs" and in the reason. The API names in the reason are already enclosed in angle brackets (e.g., \texttt{<API\_sendgrid>}).

\textbf{2.} For EACH API, evaluate the reason based on three criteria independently and provide:

   Criterion 1 Score (0-1): How well the reason aligns with the mashup requirements.
   
   Criterion 2 Score (0-1): How well the reason aligns with the API description.
   
   Criterion 3 Score (0-1): The readability of the reasoning.
   
   A concise justification for each criterion
   
\textbf{3.} Finally, provide an overall score (0 to 1) calculated as the average of all criteria  scores across ALL APIs.

If any API doesn't have a reason or any API is in the reason but not listed in “Recommended APIs", it should be given 0 for all criteria.

\textbf{\#Output Format}

Please respond in the following format:
\begin{verbatim}
{
  "api_evaluations": [
    {
      "api_name": "<API name>",
      "criterion_1_score": "<score 0-1>",
      "criterion_1_reason": "<reason>",
      "criterion_2_score": "<score 0-1>",
      "criterion_2_reason": "<reason>",
      "criterion_3_score": "<score 0-1>",
      "criterion_3_reason": "<reason>"
    }, ...
  ],
  "overall_score": "<score 0-1>"
}
\end{verbatim}
\textbf{\#Input}
          
          \textbf{Mashup}: Inflo collects day-to-day information stream like email appointment instant...
          
          \textbf{Recommended APIs}: flickr, blogger, youtube, last.fm, facebook, twitter, yahoo-mail, bing
          
          \textbf{Recommended APIs Description}: flickr: flickr api used retrieve photo...; blogger: blogger data api allows client application view...
        
        \textbf{Reason}: \texttt{<API\_flickr>} : To retrieve photo from Flickr's... \texttt{<API\_blogger>} : To fetch blog posts...
      \\\midrule
      \textbf{Answer} 
        & \begin{verbatim}
{
  "api_evaluations": [
    {
      "api_name": "flickr",
      "criterion_1_score": "0.6",
      "criterion_1_reason": 
      "...partially aligns, but... ",
      "criterion_2_score": "1.0",
      "criterion_2_reason": 
      "...matches API's capabilities...",
      "criterion_3_score": "0.8",
      "criterion_3_reason": 
      "...readable, but irrelevant..."
    }, ...
  ],
  "overall_score": "0.65"
}
\end{verbatim}
      \\\bottomrule
    \end{tabular}%

\end{table}

\subsection{Overall Recommendation Performance (RQ1)}

\begin{table}[htbp]
\centering
\caption{Performance comparison of baseline models on the task of Web API recommendation.}
\begin{tabular}{lcccccc}
\toprule[1pt]
\textbf{Methods} & \textbf{P@1} & \textbf{P@3/$s$} & \textbf{R@1} & \textbf{R@3/$s$} & \textbf{N@3/$s$} & \textbf{M@3/$s$} \\
\midrule[0.5pt] 
Llama-3.3-70B\cite{grattafiori2024llama3herdmodels}& 0.545 & 0.314 & 0.382 & 0.601 & 0.580 & 0.521 \\
Qwen3-235B-A22B\cite{yang2025qwen3technicalreport}     & 0.586 & 0.301 & 0.387 & 0.518 & 0.552 & 0.510 \\
GPT-4.1 & 0.602 & 0.333 & 0.352 & 0.591 & 0.592 & 0.564 \\
MTFM\cite{wu2021mashup}  & 0.604 & 0.294 & 0.411 & 0.526 & 0.661 & 0.646 \\
CLJT\cite{10793452} & 0.368 & 0.214 & 0.170  & 0.236 & 0.304 & 0.246 \\ 
BERT-CM\cite{10489941} & 0.680 & 0.334 & 0.477 & 0.612 & 0.730 & 0.715 \\
SEHGN\cite{WangXLPWDY24}     & 0.653 & 0.332 & 0.482 & 0.661 & 0.662 & 0.613 \\
LLMAR\cite{10948476} & 0.759 & 0.645 &0.523 &0.618 & 0.782 & 0.777\\
\textbf{WAR-R1} & \textbf{0.813} & \textbf{0.767} & \textbf{0.580} & \textbf{0.709} & \textbf{0.834} & \textbf{0.827} \\

\bottomrule[1pt]
\end{tabular}

\label{tab:result}
\end{table}
\begin{table*}[t]
\centering
\caption{Performance comparison of ablation models on dual tasks.}
\setlength{\tabcolsep}{1.0mm}
\begin{tabularx}{\linewidth}{@{}l|*{6}{>{\centering\arraybackslash}X}|*{2}{>{\hsize=0.7\hsize\centering\arraybackslash}X} *{4}{>{\hsize=1.15\hsize\centering\arraybackslash}X}@{}}
\toprule[1pt]
\multirow{2}{*}{\textbf{Models}}
  & \multicolumn{6}{c}{\textbf{Recommendation}}
  & \multicolumn{6}{c}{\textbf{Reasoning}} \\
  \cmidrule{2-7} \cmidrule{8-13}
  & \textbf{P@1}
  & \textbf{P@$s$}
  & \textbf{R@1}
  & \textbf{R@$s$}
  & \textbf{N@$s$}
  & \textbf{M@$s$}
  & \textbf{RP}
  & \textbf{RR}
  & \textbf{RS(Overall)}
  & \textbf{RS(C1)}
  & \textbf{RS(C2)}
  & \textbf{RS(C3)} \\
\midrule[0.5pt]
WAR-R1(w/o GRPO) & 0.812 & 0.748 & 0.579 & 0.712 & 0.832 & 0.825 & 0.885 & 0.891 & 0.746 & 0.741 & 0.750 &  0.748\\
WAR-R1(w/o Tokens) & 0.811 & 0.661 & 0.578 & 0.726 & 0.836 & 0.827 & 0.869 & 0.873 & 0.733 & 0.727 & 0.735 & 0.737 \\
WAR-R1(w/o Reasoning) & 0.773 & 0.727 & 0.550 & 0.676 & 0.798 & 0.790 & — & — & — & — & — & — \\
\textbf{WAR-R1} & \textbf{0.813} & \textbf{0.767} & \textbf{0.580} & \textbf{0.709} & \textbf{0.834} & \textbf{0.827} & \textbf{0.908} & \textbf{0.912} & \textbf{0.768} & \textbf{0.772} & \textbf{0.766} & \textbf{0.766} \\

\bottomrule[1pt]
\end{tabularx}
\label{tab:result_abl}
\vspace{-1.0em}
\end{table*}

Table~\ref{tab:result} reports the recommendation performance of all methods. WAR-R1 consistently outperforms all baselines across all metrics.
The LLMs Llama-3.3-70B, Qwen3-235B-A22B, and GPT-4.1 occasionally produce inappropriate recommendations due to their lack of domain-specific knowledge. MTFM, CLJT, BERT-CM, and SEHGN show varying levels of effectiveness but consistently underperform compared to WAR-R1. Although LLMAR remains efficient compared to earlier models, it lacks both special token constraints for structured API generation and semantic reasoning capabilities.
Compared with the SOTA baseline LLMAR, WAR-R1 achieves improvements of 7.11\% in Precision@1 and 10.89\% in Recall@1, and also demonstrates superior ranking quality in terms of NDCG and MAP.
Under greedy decoding, WAR-R1 generates an average of 1.79 APIs per output, with 52.22\% of outputs fully matching the ground truth, including one mashup with 31 ground-truth APIs. Built on the pretrained TinyLlama, WAR-R1 effectively captures long-range dependencies and subtle semantic connections between mashup requirements and API descriptions. 
Traditional and hybrid deep learning methods perform competitively but are limited by fixed top-$N$ strategies. In contrast, WAR-R1 adaptively adjusts the number of recommended APIs, avoiding redundant suggestions for simple mashups while maintaining coverage for complex ones.

\subsection{Evaluation of Semantic Reasoning (RQ2)}
WAR‑R1 consistently generates high‑quality semantic reasons for its API recommendations. RS is averaged over three independent evaluations of GPT-4.1 to ensure stability. Across all recommended APIs, 87.88\% are accompanied by a corresponding rationale. Table~\ref{tab:result_abl} summarizes the detailed metrics of WAR-R1 computed over each mashup instance. For RP, 84.75\% of instances achieve the maximum score of 1, yielding an average RP of 0.908. For RR, 85.48\% of instances score a perfect 1, with an average RR of 0.912. With respect to RS, the proportion of instances scoring between 0.9 and 1.0 stands at 64.24\% under GPT-4.1, with an overall average RS of 0.768. WAR-R1 achieves scores of 0.772, 0.766, and 0.766 for Criterion 1 (C1: Mashup Alignment), Criterion 2 (C2: API Alignment), and Criterion 3 (C3: Readability), respectively. 
These results demonstrate that WAR-R1 produces coherent, semantically grounded explanations rather than post hoc or generic descriptions.

\subsection{Ablation Studies (RQ3–RQ5)}
\subsubsection{RQ3: Effectiveness of the second stage training}
The second-stage GRPO training improves both recommendation accuracy and reasoning quality.
We compare the full WAR‑R1 model against an ablated variant, WAR-R1(w/o GRPO), which omits the second stage GRPO training. WAR-R1(w/o GRPO) recommends an average of 1.95 APIs per mashup with 83.32\% of these accompanied by rationales, whereas the full WAR‑R1 model issues 1.79 recommendations per mashup, 87.88\% of which include rationales. Table~\ref{tab:result_abl} reports the core performance metrics on the Web API recommendation task. By filtering out recommendations without rationales, Precision@$s$ improves by 2.54\%. Likewise, RP and RR increase by 2.60\% and 2.36\% respectively, and the proportion of instances achieving both RP=1 and RR=1 rises from 80.07\% to 83.77\%. Finally, the overall Reasoning Score increases by 2.95\%. These results confirm that the second‐stage training enhances both the accuracy of recommendation and the quality of semantic reasoning.

\subsubsection{RQ4: Effectiveness of the special tokens}
The inclusion of special tokens markedly enhances control over API and rationale generation, boosting overall performance.
By removing the special tokens \texttt{<API\_start>}, \texttt{<API\_stop>}, \texttt{<REASON\_start>} and \texttt{<REASON\_stop>}, we derive WAR-R1(w/o Tokens). Without the  constraint of these special tokens, WAR-R1(w/o Tokens) recommends an average of 2.57 APIs per mashup, but only 77.95\% of them include an accompanying rationale. Consequently, although WAR-R1(w/o Tokens) achieves a slightly higher R@$s$ than WAR‑R1, its Precision, RP, RR and RS all decline. Specifically, as shown in Table~\ref{tab:result_abl}, P@$s$ drops by 13.82\%, RP by 4.30\%, RR by 4.28\%, overall RS by 4.56\%. 
These special tokens enable WAR‑R1 to better segment and control the initiation and termination of both API recommendation and semantic reasoning. 
\subsubsection{RQ5: Impact of semantic reasoning on API recommendation}
Integrating reasoning greatly elevates API recommendation effectiveness across all metrics.
 To assess how the inclusion of reasoning affects recommendation performance, we introduce an ablation variant, WAR‑R1(w/o Reasoning), which omits the reasoning component. WAR‑R1(w/o Reasoning) is trained in two stages to ensure a fair comparison. In the first stage, the model is presented with the same API function descriptions utilized by the reasoning component of WAR‑R1. In the second stage, we fine‐tune WAR‑R1(w/o Reasoning) solely on the recommendation task. As reported in Table \ref{tab:result_abl}, integrating reasoning into WAR-R1 yields notable improvements over WAR-R1(w/o Reasoning) with 5.50\% higher P@$s$, 4.88\% higher R@$s$, 4.51\% higher N@$s$, and 4.68\% higher M@$s$. To quantify the impact of reasoning on recommendation quality, we separately evaluate metrics for recommendations accompanied by reasons versus those without. Recommendations with reasons (P@s=0.796, R@s=0.706, N@s=0.820, M@s=0.842) outperform those without (P@s=0.432, R@s=0.240, N@s=0.439, M@s=0.448) by at least 84.26\% across all metrics. These findings demonstrate that requiring the model to justify each suggestion adds rigor: semantically defensible API choices are favored, reducing spurious outputs and improving reliability.  
 

\begin{table}[t]
\caption{Computational cost of WAR-R1 and the baselines}\label{tab:cost} 
\centering
\begin{tabular}{l|c|c|c|c}
\toprule 
\textbf{Methods} & \textbf{Parameters}& \textbf{Training Time} & \textbf{Inference Time} & \textbf{GPU Usage} \\
\midrule
MTFM & 17.9 M& 3.3 m & 0.6 ms & 0.6 GB \\
CLJT &110.8 M & 5.2 m & 0.4 ms & 2.3 GB \\
BERT-CM &110.7 M & 0.2 h & 6.5 ms & 1.8 GB \\
SEHGN &3.7 M & 0.4 h & 3.7 ms & 1.1 GB \\
LLMAR &3.2 B & 1.2 h & 0.4 s & 5.5 GB \\
\textbf{WAR-R1} &\textbf{1.1 B}& \textbf{5.6 h} & \textbf{2.9 s} & \textbf{2.8 GB} \\
\bottomrule
\end{tabular}
\end{table}

\begin{table*}[tb]
  \caption{Case Study Analyses of WAR-R1.}
  \label{tab:case_study_1}
  \begin{tabularx}{\textwidth}{@{}p{0.2\textwidth} X@{}}
    \toprule
    \multicolumn{2}{@{}l}{\textbf{Mashup: betterhome }} \\
    \textbf{Target APIs}: & zillow, walk-score, google-maps, geocoder, trulia, factual-v3, yelp-fusion, socrata-open-data \\
    \textbf{Description}: & Coined "eharmony home" (for) finding a better home, it integrates different API calls to match users' ideal home, currently mashups specifically in San Francisco, CA. \\
    \midrule
    \textbf{WAR-R1} & \\
    \textbf{Generated APIs:} &\textbf{zillow}, \textbf{walk-score}, \textbf{google-maps}, \textbf{geocoder}, \textbf{trulia}, \textbf{factual-v3}, \textbf{yelp-fusion}, \textbf{socrata-open-data} \\
    \textbf{Generated Reason}: & \textbf{zillow} : Provides detailed property listings, home valuations (Zestimate), historical sale data, and neighborhood statistics essential for matching users with optimal homes based on real estate criteria. \textbf{walk-score} : Calculates walkability scores for addresses, enabling evaluation of proximity to amenities—critical for assessing urban infrastructure in locations like San Francisco. \textbf{google-maps} : Offers mapping, geolocation, and visualization (e.g., embedded maps, Street View) to display home locations, nearby points of interest, and spatial context. \textbf{geocoder} : Converts addresses to precise latitude/longitude coordinates (e.g., for Bay Area residences), required by walk-score and google-maps to analyze location-based metrics. \textbf{trulia} : Supplements real estate data with local market trends, pricing insights, and neighborhood information, enriching property comparisons. \textbf{factual-v3} : Delivers structured local data (e.g., points of interest, demographics) to profile areas beyond properties, such as neighborliness or school quality. \textbf{yelp-fusion} : Retrieves nearby business ratings (e.g., restaurants, shops) via location/phone, highlighting neighborhood vibrancy and amenories critical for home satisfaction. \textbf{socrata-open-data}: Accesses civic datasets (e.g., crime rates, transit) specific to San Francisco, providing hyperlocal insights for informed home-matching decisions. \\
    \textbf{Reasoning Score}: & 
    Overall : 0.986 , Criterion 1: 0.986, Criterion 2: 0.986, Criterion 3: 0.986 

    \\
     \midrule
      \multicolumn{2}{@{}l}{\textbf{Mashup: map-your-list}} \\
    \textbf{Target APIs}: & google-maps, google-geocoding, scribble-maps\\
    \textbf{Description}: & Convert list to map, add annotations, share, and export map in a variety of formats. \\

    \midrule

    \textbf{WAR-R1} & \\

    \textbf{Generated APIs:} &\textbf{google-maps}, \textbf{google-geocoding}, \textbf{scribble-maps}\\
    \textbf{Generated Reason}: & \textbf{google-geocoding} is used to convert the input address list into geographic coordinates, enabling the placement of locations on the map.  \textbf{google-maps} provides the foundational map display and supports exporting static maps in multiple formats (e.g., PNG, JPEG) for sharing. \textbf{scribble-maps} allows adding custom annotations (drawings, markers) to the map and facilitates direct sharing/exporting of the edited map in diverse formats (PDF, image, etc.) via its specialized drawing and collaboration features. \\
    \textbf{Reasoning Score}: &  Overall : 0.989 , Criterion 1: 1.000, Criterion 2: 1.000, Criterion 3: 0.967\\
     \midrule
   \multicolumn{2}{@{}l}{\textbf{Mashup: minguhl-bmx}} \\
    \textbf{Target APIs}: & amazon-product-advertising, ebay\\
    \textbf{Description}: & Browse detailed specs, bmx bike components; compare offers on Amazon, eBay; comment on and review products; watch bmx videos; get bmx news on top bmx sites—all in one place. Thanks, bmx Yahoo Pipe. \\

    \midrule

    \textbf{WAR-R1} & \\

    \textbf{Generated APIs:} &\textbf{amazon-product-advertising}, \textbf{ebay}, youtube \\
    \textbf{Generated Reason}: &The mashup uses \textbf{amazon-product-advertising}  to retrieve detailed Bicycle Manufacturing (Bike) components, such as frames, parts, or accessories directly from Amazon’s catalog, enabling users to discover and compile shopping deals. It leverages \textbf{ebay} (via its Finding API) to search for used Bike equipment listings on eBay, accessing transactional data and crowd-sourced reviews for informed purchasing decisions. youtube is integrated to fetch and display relevant Video content about Bikes, including reviews, tutorials, or promotional clips, utilizing its RESTful API to stream multimedia resources without redirecting users externally. \\
    \textbf{Reasoning Score}: & Overall : 1.000 , Criterion 1: 1.000, Criterion 2: 1.000, Criterion 3: 1.000\\
        \midrule
       \multicolumn{2}{@{}l}{\textbf{Mashup:wegooo}} \\
    \textbf{Target APIs}: & flickr, eventful, google-maps, zvents, last.fm, openstreetmap, geonames, weatherbug, facebook, twitter, thisnext, wikipedia, jigsaw, social-share-count, ip-location \\

    \textbf{Description}: & Wegooo wants to help (people) live in a better city. The city wants to put a finger on its pulse, to help understand the spirit of the people who inhabit it, and enjoy what was formerly called "enjoy everywhere". \\
    \midrule

    \textbf{WAR-R1 } & \\

    \textbf{Generated APIs:} & \textbf{flickr}, \textbf{eventful}, \textbf{google-maps}, \underline{\textbf{zvents}}, \textbf{last.fm}, \textbf{openstreetmap}, \textbf{geonames}, \textbf{weatherbug}, \textbf{facebook}, \textbf{twitter}, \underline{\textbf{thisnext}}, \underline{\textbf{wikipedia}} \\

    \textbf{Generated Reason}: & Based on the mashup description "wegooo wants to help living better city", here's why each API is used: \textbf{flickr}: To retrieve photos showcasing urban environments, landmarks, and atmosphere, visually capturing the "spirit" of cities.  \textbf{eventful} : To find local events (concerts, sports, rallies) happening in the searched city, reflecting what people are enjoying ("enjoy") happens there. \textbf{google-maps} : To provide detailed maps showing locations, routes, and navigation options, helping users explore areas within a givencity.  \underline{\underline{upcoming.org}}: To discover nearby restaurants, bars, cafes, and points interest, adding food quality-of-life details to the city discovery experience.  \textbf{last.fm} : To recommend music festivals, concerts, or popular cultural experiences tied to venues/artists in the city. \textbf{openstreetmap} : To offer historical context about neighborhoods, landmarks, and culture by linking places to their medieval past. \textbf{geonames}: To supply coordinates, map entire world, and identify points interesting enough to warrant inclusion ("whereness"). \textbf{weatherbug}: To provide real-time weather conditions and forecasts ensuring optimal planning sessions around outdoor activities.  \textbf{facebook}: To enable social sharing of explored cities, shares check-ins, and allows accessing friend networks for community building. \textbf{twitter} : To integrate real-time updates on artists touring the city, highlighting local entertainment opportunities.  \underline{\underline{bandsintown}}: To display geolocated user posts offering insights into daily Life InTheCity.com visitors.  \\
     \textbf{Reasoning Score}: & Overall : 0.619 , Criterion 1: 0.607, Criterion 2: 0.607, Criterion 3: 0.643\\

    \bottomrule
\multicolumn{2}{p{\textwidth}}{$^{\mathrm{a}}$Correctly recommended APIs are shown in \textbf{bold}; incorrectly recommended APIs are shown in plain text; APIs recommended but not mentioned in the reason are \underline{underlined}; APIs mentioned in the reason but not recommended are  (\underline{\underline{double underlined}}).} 
  \end{tabularx}
\end{table*}

\subsection{Computational Cost Analyses (RQ6)}
We compare WAR-R1 with various baseline approaches across multiple dimensions, including total parameters of the model, training time to reach the model's best performance, inference time per query, and inference GPU memory usage, as shown in Table~\ref{tab:cost}. 
Due to the additional reasoning component, WAR-R1 takes longer to train and perform inference. 
The first stage training requires 1.1 hours, and the second stage requires 4.5 hours. 
Without using LoRA, it could take more than 22 hours to train.
Although the training and inference times are longer, WAR-R1 remains practical, requiring only 2.8 GB of GPU memory at inference time, while providing explainable Web API recommendation results.

\subsection{Further Insights into Reasoning Score}
\subsubsection{Alignment with human preference and metrics}
To validate the reliability of our automated evaluation, we recruited 52 participants with knowledge of Web services or Web applications from the Prolific \cite{article} platform to annotate 600 records. The human annotators evaluated each record based on the same three criteria used by GPT-4.1. The average human scores are 0.779, 0.763, and 0.784 for Criterion 1, Criterion 2, and Criterion 3, respectively. To measure the agreement between human annotations and GPT-4.1 scores, we compute both Pearson correlation coefficient ($r$) and Spearman's rank correlation coefficient ($\rho_s$). 
$r$ captures the linear relationship between the two sets of scores, while $\rho_s$ measures the monotonic relationship based on rank ordering. 
The overall alignment achieves $r$ = 0.806 and $\rho_s$ = 0.687. At the criterion level, C1 obtains $r$ = 0.749 and $\rho_s$ = 0.623, showing the strongest agreement, as human annotators and GPT-4.1 share a consistent understanding of requirement relevance. C2 obtains $r$ = 0.748 and $\rho_s$ = 0.562, where the lower rank correlation likely reflects differing interpretations with technical API terminology between human annotators and GPT-4.1. C3 obtains $r$ = 0.715 and $\rho_s$ = 0.584, as human annotators tend to prioritize comprehensibility whereas GPT-4.1 additionally penalizes verbosity and formatting inconsistencies. The overall alignment is stronger than individual criteria, as discrepancies across dimensions tend to compensate when aggregated. Moreover, the correlations of RS with RR and RP yield $r$ = 0.794 and $r$ = 0.812 respectively, further validating RS as an automated assessment metric aligning with human preference. 
 
 \subsubsection{Annotated dataset evaluated by Reasoning Score}
The annotated dataset achieves an average RS of 0.973, with individual criterion scores of 0.970 for Criterion 1, 0.977 for Criterion 2, and 0.973 for Criterion 3. Notably, over 94.13\% of records achieve scores exceeding 0.9 across all criteria.
Ideally, the annotated dataset, serving as ground truth, should receive a full score of 1.0. However, GPT-4.1 demonstrates considerable rigor as an evaluator, occasionally assigning scores of 0.8 or 0.9 rather than perfect scores. For C1, it provides justifications such as: ``The reason states use for user authentication, which is relevant but not core mashup functionality," emphasizing whether the reasoning explicitly connects API functions to the core requirements of the mashup. For C2, typical explanations include: ``The reason mostly aligns with the API description, though it simplifies the API's scope," focusing on whether the reasoning accurately reflects the API's documented capabilities and scope. For C3, representative rationales are: ``The explanation could be slightly more specific about how Burstn integrates into the search functionality" or ``The reason is clear but slightly verbose," suggesting that GPT-4.1 values both specificity and conciseness in the explanations.
Despite these strict evaluation standards, the annotated dataset by DeepSeek-R1 achieves consistently high scores, confirming its reliability.
\section{Case Study}
Table \ref{tab:case_study_1} presents four mashups from ProgrammableWeb, illustrating WAR-R1's ability to recover key services and provide intuitive explanations. In ``betterhome" and ``map-your-list",  WAR-R1 correctly identifies all target APIs and produces coherent, semantically grounded explanations, achieving RS scores of 0.986 and 0.989. GPT-4.1 generally confirms strong alignment across all three criteria, while identifying minor areas for improvement, for instance, noting that the explanation for factual-v3 would benefit from more mashup-specific details.
For ``minguhl-bmx", the model correctly recommends amazon-product-advertising and ebay. Although youtube is not used in the original mashup, WAR-R1's explanation highlights its utility for streaming product demonstrations, indicating that beyond functional alignment, real-world API selection may also involve practical considerations.
For ``wegooo," a complex mashup invoking 15 APIs, WAR-R1 retrieves 12 target services. This case reveals partial misalignment between recommendation and reasoning: three APIs (zvents, thisnext, wikipedia) lack explanations, while two (upcoming.org, bandsintown) appear only in the reasoning text. GPT-4.1 assigns scores of 0 to all misaligned APIs, demonstrating its ability to enforce structural consistency. Furthermore, GPT-4.1 detects a functional misrepresentation where WAR-R1 incorrectly characterizes openstreetmap as providing historical context rather than its actual mapping functionality. GPT-4.1 assigns lower scores for Criterion 1 and Criterion 2 while still awarding a full score for Criterion 3, showing that the evaluation framework can distinguish factual accuracy from writing quality at a fine-grained level.  
Overall, the case study demonstrates that WAR-R1 produces focused and interpretable recommendations whose output size naturally adapts to mashup complexity while also exposing realistic challenges associated with free-form semantic reasoning.

\section{Discussion and Limitations}
\label{sec:discuss}
The experimental results indicate that WAR-R1 can generate Web API recommendations along with high-quality reasoning as justification. 
Despite these promising results, several limitations remain. First, the current reasoning focuses primarily on functional aspects, while real-world API selection often involves additional factors such as QoS, pricing, and reliability. Incorporating datasets annotated by experts in mashup development that capture these practical considerations could further improve recommendation quality. Second, the reasoning metrics RR and RP have not yet reached perfect scores. We explored alternative architectures such as reason before recommend, however, none of these approaches yielded better performance compared to our current design, potentially due to the limited parameter capacity of the base model. 
Although evaluated on Web API recommendation, the proposed framework is applicable to other recommendation scenarios that require adaptive set prediction and explainability, such as service composition and software component recommendation.
\section{Conclusion}
\label{sec:conclu}
In this study, we propose WAR-R1, an explainable Web API recommendation framework that integrates semantic reasoning with adaptive, variable-cardinality recommendation. By introducing structured generation with control tokens and a two-stage training strategy combining supervised fine-tuning and GRPO-based reinforcement learning, WAR-R1 jointly optimizes recommendation accuracy and explanation quality.
Experiments demonstrate WAR-R1 outperforms SOTA baselines while producing coherent and semantically grounded explanations. 
Ablation studies confirm the effectiveness of reinforcement learning, structured generation, and reasoning integration.
This work shows that jointly modeling \textit{what to recommend and why to recommend it} leads to more accurate, transparent, and developer-friendly Web API recommendation. We believe WAR-R1 represents a practical step toward trustworthy intelligent services and provides a basis for future research on explainable and adaptive recommendation systems.

\clearpage

\bibliography{ICDCS2026/cite.bib}
\end{document}